\documentclass[twocolumn,twoside,slac_two]{revtex4}
\usepackage{graphicx}
\usepackage{fancyhdr}
\begin{document}

\title{Testing the Friedmannian magnitude-redshift relation with SNIa data}

%

\author{M. N. C\'el\'erier}
\affiliation{Laboratoire Univers et TH\'eories (LUTH) Observatoire 
de Paris-Meudon, 5 place Jules Janssen, 92195 Meudon, FRANCE}

\begin{abstract}

Standard cosmology is constructed upon the (generally implicit) assumption of the ``large scale'' homogeneity of our Universe. Now, structures are observed at scales which become larger and larger as the observational distances increase. However, the homogeneous Friedmann-Lemaitre-Robertson-Walker model remains a cosmological paradigm, and Friedmannian relations are usually used to work out the model, implicitly assuming that the homogeneity assumption is valid at the studied scales. This has been the way the concordance ($\Lambda$CDM) model came out from the analyses of the SNIa data. Since this model implies the validity of the Friedmannian magnitude-redshift relation at the range of redshifts spaned by the supernovae surveys, it has been proposed some years ago a very simple test of this possible validity, provided the SNIa should be confirmed as standard candles. Since we can hope that the redshifts of the supernovae which will be observed in the years to come will allow such a test to be performed, it would be interesting to find inside the SNIa community some collaborators ready to perform the test. 
\end{abstract}

\maketitle

\thispagestyle{fancy}


\section{The three observational pilars of the standard Hot Big-Bang model}

The standard Hot Big-Bang model is based upon three observational pilars, of which the most common interpretations have been thoroughly discussed in a review paper published in 2003 by L\'opez-Corredoira~\cite{MLC03}. We refer here to this author, 
even if some of the possible caveats proposed in his work seem 
too weak to really shake the actual robustness of these pilars, 
as far as our current scientific knowledge is involved.
\begin{enumerate}

 \item The increasing with distance galaxy redshifts are commonly 
interpreted as due to a recession of the galaxies implying an  expanding Universe. This interpretation seems to be the robustest available upon the market even if some authors have proposed other ones.
 
 \item The measured abundance of the light elements is generally explained in the framework of the standard Big Bang primordial nucleosynthesis. Alternatives can be found in the literature, although this model might be considered as the best proposed up to now, besides some possible problems that might be or not solved in the future.
 
 \item The cosmic microwave background (CMB) is mostly believed to be a relic of the early stage of the Universe, when the photon decoupled from the high energy primordial soup, that is observed as a background with a perfect blackbody spectrum. By now, nobody has proposed a satisfactory alternative scenario and therefore the standard interpretation remains the best one. However, L\'opez-Corredoira~\cite{MLC03} points out, for example, a possible problem arising from the high energy cosmic rays. These are currently measured with energies up to 3.2 $10^{20}$ eV, which is, in the framework of standard physics (another explanation has been proposed in the framework of Scale Relativity~\cite{LN03}) beyond the theoretical energy limit for particles traveling for distances such as those separating powerful galaxies, because of their interactions with the CMB.  Considering the robustness of the other observations in favor of the standard interpretation of the CMB in the Hot Big-Bang model, it seems more clever to consider that this problem pertains to particle physics rather than to cosmology.
 
  \end{enumerate}

\section{The``concordance'' $\Lambda$CDM model}

The current program of a major part of the cosmological community is to determine a set of cosmological ``constants'' in the framework of a Friedmann-Lema\^itre-Robertson-Walker (FLRW) model of Universe incorporating an inflationary scenario.

\begin{itemize}
  \item The Hubble Space Telescope (HST) provides observations allowing an estimate of the Hubble constant $H_0$.
  \item The deuterium abondance measured in Lyman $\alpha$ clouds gives a value for $\Omega_b$.
  \item The abundance of rich clusters in large scale structures  leads to an estimate for $\sigma_8$.
  \item Large scale weak lensing is a probe for $\Omega_M$ and $w$.
  \item The magnitude-redshift relation issuing from the SNIa observations provides ranges for $H_0$, $\Omega_M$ and $\Omega_{\Lambda}$
  \item The CMB $C_l$ curve is used to put constraints on 6 to 12 cosmological "constants", the number of constants depending on the cosmological assumptions retained.
  \end{itemize}
  
  A cross-correlation of all these results, and others that cannot be cited here, an exhaustive review of the cosmological field being beyond the scope of this contribution, provides what is known as the concordance or $\Lambda$CDM model, with the following values of the main cosmological``constants'' pertaining to this model (see, e.g.,~\cite{AB03}) : \\
  $H_0 \sim 70$km/s/Mpc, $\Omega_b \sim 0.02 h^{-2}$, $\Omega_0 \sim 1$, $\Omega_M \sim 0.3$, $\Omega_{\Lambda} \sim 0.7$, $n\sim 1$, $\tau \sim 0$, etc.

\subsection{Priors to the ``concordance'' model:}

However, the ``concordance'' model is built with priors pertaining to the ``standard'' paradigm of cosmology and which are:
\begin{itemize}
  \item The FLRW metric of the Universe. \\
  Since all isotropic, homogeneous, uniformly 
  expanding universes exhibit a FLRW metric~\cite{ML03}, their geometry can be 
  caracterized by a finite number of cosmological ``constants''. \\
  It is worth stressing here that the FLRW solution implies, as an implicit assumption, that global solutions to Einstein's equations can be physical, although we know that General Relativity is an essentially local theory. A lot of work can be found in the literature (for a review see, e.g.,~\cite{AK97})  which discuss the validity of such an approach. In the FLRW case, the stress-energy tensor is averaged over a scale which is generally neither precised nor probed. However, there is currently no available mathematical definition of a tensor average. Averaging is only mathematically valid for scalars. Therefore, Friedmannian relations must be used with great care and should be considered as approximations and systematically validated for the scale range to which they are applied.
  
  \item The inflation assumption. \\
  It implies a power-law primordial spectrum of the density perturbations presumably generated during an inflationary phase and assumed for the analysis of the CMB temperature fluctuations.   Another rather generic prediction of this paradigm is a strict spatial flatness of the Universe, implying $\Omega_0 = 1$. The current error bars on $\Omega_0$ ($\Omega_0 = 1.02 \pm 0.02$~\cite{DNS03}) are still too large to test this prediction, which might be more easy to discard than to validate, since the prediction is $\Omega_0 =$ (strictly) $1$. 
  \end{itemize}
  
  \subsection{Other theoretical alternatives}
Other theoretical alternatives have been proposed by different authors. The reader will find below some among the most famous examples:
  \begin{itemize}
  \item Gaussian adiabatic fluctuations versus isocurvature
  \item $\Lambda$CDM with a cosmological constant versus quintessence
  \item vanishing versus non vanishing amount of gravitational waves
  \item vanishing versus non vanishing cosmic string component
  \item etc.
  \end{itemize}

\section{The homogeneity assumption}

 ``Large scale spatial homogeneity'' is generally assumed without any precision as to the scale at which the Universe is supposed to become homogeneous. However, large scale inhomogeneities can yield a geometry noticeably different from the one infered from 
  the homogeneous assumption, therefore invalidating the use of Friedmannian expressions. Inhomogeneous solutions to Einstein's equations generally involve cosmological functions of the coordinates rather than cosmological constants as in FLRW solutions (see, e.g., for a mere example,~\cite{MNC00}). \\

As a consequence, the expressions or relations retained to analyse data collected at any distances must be carefully used, in particular those assuming the ``constance'' of the cosmological ``constants''. \\
 
 But these expressions can be used to test some of the 
 features of the geometry of the Universe. This is, in particular, the case of the FLRW expression for the luminosity distance as a function of redshift and of the three``constants'' $H_0$, $\Omega_M$ and $\Omega_{\Lambda}$, which has been initially retained to analyse the data of the SNIa~\cite{AGR98,SP99}. \\

 \section{Use of the magnitude redshift-relation to test some features of the geometry of the Universe}

  The FLRW model implies the use of a relation for the luminosity distance $D_L$ as a function of the four parameters,  $z$, $H_0$, $\Omega_M$ and $\Omega_{\Lambda}$. This is only valid if the FLRW picture is confirmed as a good approximation for 
  all the spaned values of $z$. The relation is the following:
  
  \begin{eqnarray}
D_L={c(1+z)\over {H_0\sqrt{|\kappa |}}} \,\, {\cal S}&\biggl(&\sqrt
{|\kappa |}\int^z_0[(1+z')^2(1+\Omega _M z') \nonumber \\
&&-z'(2+z')\Omega _\Lambda]^{-{1\over 2}}
dz'\biggr) ,
\end{eqnarray}
\begin{tabular}{cll}
for $\Omega _M + \Omega _\Lambda > 1$ & ${\cal S}=\sin$ & and 
$\kappa = 1-\Omega _M-\Omega _\Lambda$ \\
for $\Omega _M + \Omega _\Lambda < 1$ & ${\cal S}=\sinh$ & and 
$\kappa = 1-\Omega _M-\Omega _\Lambda$ \\
for $\Omega _M + \Omega _\Lambda = 1$ & ${\cal S}=$ I & and 
$\kappa = 1$ . \\
\end{tabular}

   The apparent magnitude of a source at a given $z$ writes
   \begin{equation}
m=M+5 \log D_L(z;H_0, \Omega _M,\Omega _\Lambda)+25 .
   \end{equation}
    
   The proposed method, inspired from Goobar and Perlmutter, 1995~\cite{AG95} uses this expression to draw contours of constant apparent magnitude on the $\Omega _\Lambda$-versus-$\Omega _M$ plane for different fixed redshifts. One first selects, from 
   the SNIa catalogs, three fair samples at sufficiently different redshifts, $z_1<z_2<z_3$. Sets of rather low redshifts can be used to try to find out the scale of a 
   possible transition to homogeneity, usually expected above 
   some hundred Mpc. For the higher reached $z_3$, this transition, if not found at lower redshifts, may also be put to the test. Note that the errors allowed for small redshift samples used for this kind of analysis would be lower 
   than for high redshift ones. \\
   
\begin{figure}
\includegraphics[width=65mm]{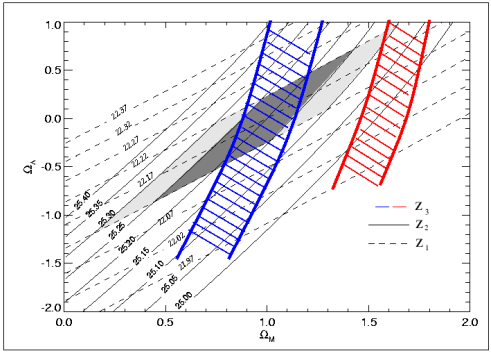}
\caption{Contours of constant $m$ in the $\Omega_M$-$\Omega_{\Lambda}$ plane.}
\label{Fig. 1}
\end{figure}
  
The uncertainties in the measurements 
   define three strips between two contour lines for each allowed range of $\Omega _M$ and $\Omega _\Lambda$. The two first ones 
   ($z \sim z_1$, $z \sim z_2$) cross, thus defining a possible allowed region (dashed rhombus in \ref{Fig. 1}). The third strip ($z \sim z_3$) does or does not cross this rhombus. 
   
\subsection{Possible missing of the rhombus}
If the data produce the red case in Figure 1, where the third strip does not cross the rhombus, the FLRW magnitude-redshift relation might be discarded, at least at the spaned redshifts. \\

As a consequence, one (or more) of the assumptions retained to derive this relation must be removed: either homogeneity, or the constance of the 
Cosmological Constant, or the universe expansion (but this 
last seems the robuster one, being one of the three pilars of the standard model of universe, see above Section 1) etc. \\
 
\subsection{ Possible crossing of the rhombus}
If the data yield the blue strip case of Figure 1, this  would imply a validation of the FLRW magnitude-redshift relation at the spaned redshifts. \\
  
As a consequence, this would be rather good support for the  homogeneity assumption, for the use of this method to obtain the value of  $\Omega_{\Lambda}$ and for its interpretation as a real Cosmological Constant versus quintessence. \\                                               
  Note that the ``concordance'' value $\Omega_{\Lambda}\sim 0.7$ corresponds to a theoretical prediction of the value of the Cosmological Constant in the framework of the scale relativity theory~\cite{LN03,LN93}. \\
 
 But one should remain careful since it has been shown that inhomogeneous cosmological models, with or without a non-zero cosmological constant, can mimic FLRW ones with a different value for $\Lambda$~\cite{MNC00,HKS92}.
   
\section{Conclusion}

\begin{itemize}
 \item The current ``concordance'' model of the Universe is a very interesting product of crossed analyses of the large amount of currently available cosmological data.
 \item We must however remain aware of its priors, mostly implicitly assumed to derive the geometry and the dynamics of our Universe with those methods.
 \item The Friedmannian magnitude-redshift 
 relation can be used as a test, using fair samples of SNIa data at different redshifts to: i) determine the redshift 
 value of the possible transition from inhomogeneity to 
 homogeneity; ii) test the Cosmological Constant 
 versus quintessence or other hypotheses; iii) and more generally  
 the FLRW model at the spaned redshifts. 
 \item It could be also an independent way of testing the validity of the analysis of the fractal dimension of the galaxy distribution currently in process~\cite{MNC05}, by comparing the values for this transition obtained by both methods.
 \end{itemize}
 
 One must however remain aware of the high degeneracy of the cosmological exercise. When dealing with only one Universe, many different answers can be proposed to a given problem. The application of Occam's razor to discriminate between a number of abundant ideas, some of them very ingenious and appealing, is the usual and more economic way to obtain at least an elegant, at most ``the'' right solution. This is the reason why the proposed test retains a set of very simple assumptions in the framework of Einstein's General Relativity. Of course, other interpretations than those proposed in Section 4 can be put forward to explain the results of the test. The high degeneracy of the cosmological problem, here stressed once again, alows each reader to choose among the possible explanations the one which corresponds to his (her) own taste, provided not too many epicycles would be needed to construct his (her) model of Universe (do not forget Occam's razor).
 
\section{Prospects}

To complete the proposed test we need:
 \begin{itemize}
 \item A validation of the SNIa as accurate standard candles, actively in progress inside the SNIa community ~\cite{PN02,AGR04}.
 \item The largest catalogs of photo-spectroscopic measurements of SNIa from very low (needed for calibration) to sufficiently high redshifts that can or will be found in the on-going or planed surveys: the Nearby Supernovae Factory (SNfactory), the Supernova Legacy Survey(SNLS), Equation of State: SupErNovae trace Cosmic Expansion (ESSENCE), the Supernova/Acceleration Probe (SNAP), etc.
 \item A collaboration with the SNIa teams to use the more recent data to perform a statistically robust analysis.
 \end{itemize}

\end{document}